\documentclass[pre,twocolumn,aps]{revtex4}
\newcommand{\bec}[1]{\mbox{\boldmath $ #1$}}
\usepackage{graphicx}
\begin{document}
\bigskip
\bigskip
\title{\bf Small-scale magnetic buoyancy and magnetic pumping effects in a
turbulent convection}
\author{IGOR ROGACHEVSKII}
\email{gary@bgu.ac.il}
\author{NATHAN KLEEORIN}
\affiliation{\rm Department of Mechanical Engineering, The
Ben-Gurion University of the Negev, POB 653, Beer-Sheva 84105,
Israel}
\date{Received  9 January 2006; in final form 15 May 2006}

\begin{abstract}
We determine the nonlinear drift velocities of the mean magnetic
field and nonlinear turbulent magnetic diffusion in a turbulent
convection. We show that the nonlinear drift velocities are caused
by the three kinds of the inhomogeneities, i.e., inhomogeneous
turbulence; the nonuniform fluid density and the nonuniform
turbulent heat flux. The inhomogeneous turbulence results in the
well-known turbulent diamagnetic and paramagnetic velocities. The
nonlinear drift velocities of the mean magnetic field cause the
small-scale magnetic buoyancy and magnetic pumping effects in the
turbulent convection. These phenomena are different from the
large-scale magnetic buoyancy and magnetic pumping effects which are
due to the effect of the mean magnetic field on the large-scale
density stratified fluid flow. The small-scale magnetic buoyancy and
magnetic pumping can be stronger than these large-scale effects when
the mean magnetic field is smaller than the equipartition field. We
discuss the small-scale magnetic buoyancy and magnetic pumping
effects in the context of the solar and stellar turbulent
convection. We demonstrate also that the nonlinear turbulent
magnetic diffusion in the turbulent convection is anisotropic even
for a weak mean magnetic field. In particular, it is enhanced in the
radial direction. The magnetic fluctuations due to the small-scale
dynamo increase the turbulent magnetic diffusion of the toroidal
component of the mean magnetic field, while they do not affect the
turbulent magnetic diffusion of the poloidal field.

\bigskip

\end{abstract}

\maketitle

\section*{1. Introduction}

Magnetic fields observed in astrophysical plasma are strongly
inhomogeneous (see, e.g., Moffatt 1978, Parker 1979, Krause and
R\"{a}dler 1980, Zeldovich {\it et al.} 1983, Ruzmaikin {\it et al.}
1988, Stix 1989, Roberts and Soward 1992, Kulsrud 1999, and
references therein). For instance, the sunspots and the solar active
regions are related to the strongly inhomogeneous large-scale
magnetic fields. The scales of the magnetic inhomogeneities, e.g.,
in the Sun are much smaller than the radius of the Sun and usually
much larger than the size of granules of the solar convection. One
of the mechanisms of the formation of the magnetic inhomogeneities
is associated with the magnetic buoyancy instability of stratified
continuous magnetic field (see, e.g., Parker 1966, Gilman 1970,
Priest 1982). The magnetic buoyancy instability is excited if the
scale of variations of the initial magnetic field is less than the
density stratification length. This mechanism does not include
explicitly magnetic flux tubes. On the other hand, the buoyancy of
the magnetic flux tubes as a mechanism of the formation of the
magnetic structures was studied in a number of publications (see,
e.g., Parker 1955, Spruit 1981, Spruit and van Ballegooijen 1982,
Sch\"{u}ssler {\it et al.} 1994, Moreno-Insertis {\it et al.} 1996).
Note also that the problem of the storage of magnetic fields and the
formation of flux tubes in the overshoot layer near the bottom of
the solar convective zone was investigated, e.g., by Spiegel and
Weiss (1980), Tobias {\it et al.} (2001), Tobias and Hughes (2004),
Brandenburg (2005).

Another universal mechanism of the formation of the nonuniform
distribution of magnetic flux in flows of the conducting fluid is a
magnetic flux expulsion. In particular, the expulsion of magnetic
flux from two-dimensional flows (a single vortex and a grid of
vortices) was demonstrated by Weiss (1966). In the context of solar
and stellar convection, the topological asymmetry of stationary
thermal convection plays very important role in the magnetic field
dynamics (Drobyshevski and Yuferev 1974). Fluid rises at the centers
of the convective cells and falls at their peripheries. This results
in that the ascending fluid elements (contrary to the descending
fluid elements) are disconnected from one another. This causes a
topological magnetic pumping effect allowing downward transport of
the mean horizontal magnetic field to the bottom of a cell but
impeding its upward return (Drobyshevski and Yuferev 1974, Zeldovich
{\it et al.} 1983, Galloway and Proctor 1983). The fine structure of
a sunspot is determined by the local interaction between magnetic
fields and turbulent convection near the Sun's surface. It was shown
recently by Thomas {\it et al.} (2002) that a downward pumping of
magnetic flux may cause of filamentary structures in sunspot
penumbrae. In particular, the magnetic field lines are kept
submerged outside the spot by turbulent, compressible convection,
which is dominated by strong, coherent, descending plumes.

Turbulence causes additional effects, e.g., the turbulent
diamagnetic and paramagnetic drift velocities of the mean magnetic
field (Zeldovich 1956, Krause and R\"{a}dler 1980, Vainshtein and
Kichatinov 1983, Kichatinov 1991, Kichatinov and R\"{u}diger 1992,
Kichatinov and Pipin 1993, Kleeorin and Rogachevskii 2003,
R\"{a}dler {\it et al.} 2003, Rogachevskii and Kleeorin 2004). In
particular, an inhomogeneity of the velocity fluctuations leads to a
transport of mean magnetic flux from regions with high intensity of
the velocity fluctuations (turbulent diamagnetism, see, e.g.,
Zeldovich 1956, Krause and R\"{a}dler 1980, Vainshtein and
Kichatinov 1983, Kichatinov and R\"{u}diger 1992, R\"{a}dler {\it et
al.} 2003). On the other hand, an inhomogeneity of magnetic
fluctuations due to the small-scale dynamo causes turbulent
paramagnetic velocity, i.e., the magnetic flux is pushed into
regions with high intensity of the magnetic fluctuations (Vainshtein
and Kichatinov 1983, Kichatinov 1991, R\"{a}dler {\it et al.} 2003).
Another effects are the effective drift velocities of the mean
magnetic field caused by inhomogeneities of the fluid density
(Kichatinov 1991, Kichatinov and R\"{u}diger 1992) and pressure
(Kichatinov and Pipin 1993). In a nonlinear stage of the magnetic
field evolution, inhomogeneities of the mean magnetic field
contribute to the diamagnetic or paramagnetic drift velocities
depending on the level of magnetic fluctuations due to the
small-scale dynamo and level of the mean magnetic field
(Rogachevskii and Kleeorin 2004). The diamagnetic velocity causes a
drift of the magnetic field components from the regions with a high
intensity of the mean magnetic field.

The pumping of magnetic flux in three-dimensional compressible
magnetoconvection has been studied in direct numerical simulations
by Ossendrijver {\it et al.} (2002) (see also review by Ossendrijver
2003). The resulting magnetic pumping effects are isolated in the
direct numerical simulations by calculating the turbulent
diamagnetic and paramagnetic velocities. The pumping effect in the
vertical direction is found as a predominating downward advection
with a maximum speed in the turbulent convection of about 10
percents of the turbulent velocity (Ossendrijver {\it et al.} 2002).

The turbulent diamagnetic and paramagnetic velocities were
determined analytically  in previous studies only for purely
hydrodynamic turbulence. A relation to the turbulent convection was
made in some studies (see, e.g., Kichatinov 1991, Kichatinov and
Pipin 1993) only phenomenologically, using the equation $\langle
{\bf u'}^2 \rangle \propto g \tau_0 \langle u'_z s' \rangle$ which
follows from the mixing-length theory. Here $\langle u'_z \, s'
\rangle$ is the vertical turbulent heat flux, ${\bf u}'$ and $s'$
are fluctuations of fluid velocity and entropy, ${\bf g}$ is the
acceleration of gravity and $\tau_0$ is the characteristic
correlation time of turbulent velocity field. This relationship
implies that the vertical turbulent heat flux plays a role of a
stirring force for the turbulence. However, a more sophisticated
approach implies a solution of a coupled system of dynamical
equations which includes the equations for the Reynolds stresses
$\langle u'_i \, u'_j \rangle$, the turbulent heat flux $\langle s'
\, u'_i \rangle$, the entropy fluctuations $\langle s' \, s'
\rangle$, the magnetic fluctuations $\langle b_i \, b_j \rangle$,
the cross helicity tensor $\langle b_i \, u'_j \rangle$ and $\langle
b_i \, s' \rangle$ in a turbulent convection. The latter has not
been taken into account in the previous studies of the small-scale
magnetic buoyancy and magnetic pumping effects caused by the
turbulent diamagnetic and paramagnetic drift velocities. Note that
the turbulent convection can strongly affect these phenomena.

The goal of this study is to determine the nonlinear drift
velocities of the mean magnetic field in a turbulent convection. We
demonstrate that the nonlinear drift velocities depend on the
different kinds of the inhomogeneities: (i) inhomogeneous
turbulence; (ii) the nonuniform fluid density and (iii) the
nonuniform turbulent heat flux. The inhomogeneous turbulence causes
the well-known turbulent diamagnetic and paramagnetic velocities. In
addition, the nonlinear drift velocities results in the small-scale
magnetic buoyancy and magnetic pumping in the turbulent convection.
These phenomena are different from the large-scale magnetic buoyancy
and magnetic pumping effects. The large-scale phenomena are caused
by the influence of the mean magnetic field on the large-scale fluid
flow. Our study shows that these large-scale effects are stronger
than the small-scale magnetic buoyancy and magnetic pumping only for
a strong mean magnetic field (about equipartition field). We study
the small-scale magnetic buoyancy and magnetic pumping effects in
the context of the solar and stellar turbulent convection. In
particular, we demonstrate that in the main part of the solar
convective zone the small-scale magnetic pumping effect dominates,
while near the solar surface the radial drift velocity of the weak
mean magnetic field results in the small-scale magnetic buoyancy
effect. We also investigate the anisotropic turbulent magnetic
diffusion of the mean magnetic field in the turbulent convection.

This paper is organized as follows. In section~2 we formulate the
governing equations, the assumptions, the procedure of the
derivations. In section~3 we consider the axisymmetric $\alpha
\Omega$ dynamo problem and determine the nonlinear drift velocities
of the mean magnetic field and nonlinear turbulent magnetic
diffusion in a turbulent convection. In section~4 we discuss the
small-scale magnetic buoyancy and magnetic pumping effects and made
estimates for the solar and stellar turbulent convection. Finally,
we draw conclusions in section~5. In Appendixes~A we perform a
detailed derivation of the nonlinear drift velocities of the mean
magnetic field and nonlinear turbulent magnetic diffusion in a
turbulent convection.

\section*{2. The governing equations}

In this study we investigate the small-scale magnetic buoyancy and
magnetic pumping effects in a turbulent convection. These phenomena
are determined by the nonlinear drift velocities in the nonlinear
electromotive force. In order to derive the nonlinear electromotive
force in the turbulent convection we use a mean field approach in
which the magnetic and velocity fields, and entropy are decomposed
into the mean and fluctuating parts, where the fluctuating parts
have zero mean values. We assume that there exists a separation of
scales, i.e., the maximum scale of turbulent motions $l_0$ is much
smaller then the characteristic scale $L$ of inhomogeneities of the
mean fields. We adopt here a procedure of the derivation of the
nonlinear electromotive force which was applied previously by
Rogachevskii and Kleeorin (2004) for the hydrodynamic incompressible
turbulence. Let us outline here the procedure of the derivation of
the nonlinear electromotive force for the turbulent convection (for
details, see also Appendix A). We consider a nonrotating turbulent
convection with large Rayleigh numbers and large hydrodynamic and
magnetic Reynolds numbers. The equations for fluctuations of the
fluid velocity, entropy and the magnetic field are given by
\begin{eqnarray}
{1 \over \sqrt{\rho_0}} {\partial {\bf v}({\bf x},t) \over
\partial t} &=& - \bec{\nabla} \biggl({p \over
\rho_0} \biggr) + {1 \over \sqrt{\rho_0}} \biggl[({\bf b} \cdot
\bec{\nabla}) {\bf H}
\nonumber \\
&& + ({\bf H} \cdot \bec{\nabla}){\bf b} + \, {\Lambda_\rho \over 2}
[2{\bf e}({\bf b} \cdot {\bf H})
\nonumber \\
&& - ({\bf b} \cdot {\bf e}) {\bf H}] \biggr] - {{\bf g} \over
\sqrt{\rho_0}} \, s + {\bf v}^N \,,
\label{B1} \\
{\partial {\bf b}({\bf x},t) \over \partial t} &=& ({\bf H} \cdot
\bec{\nabla}){\bf v} - ({\bf v} \cdot \bec{\nabla}) {\bf H} +
{\Lambda_\rho \over 2} [{\bf v} ({\bf H} \cdot {\bf e})
\nonumber \\
&& - \, {\bf H} ({\bf v} \cdot {\bf e})] + {\bf b}^N \,,
\label{B2} \\
{\partial s({\bf x},t) \over \partial t} &=& - {\Omega_{b}^{2} \over
g} ({\bf v} \cdot {\bf e}) + s^N \,, \label{B3}
\end{eqnarray}
where we use new variables $({\bf v}, \, s, \, {\bf H})$ for
fluctuating fields ${\bf v} = \sqrt{\rho_0} \, {\bf u}' $ and $s =
\sqrt{\rho_0} \, s'$, and also for the mean field ${\bf H} = {\bf B}
/ (\mu \, \sqrt{\rho_0})$. Here ${\bf B}$ is the mean magnetic
field, $\rho_0$ is the fluid density, $\mu$ is the magnetic
permeability of the fluid, ${\bf e}$ is the vertical unit vector,
$\Omega_{b}^{2} = - {\bf g} {\bf \cdot} \bec{\nabla} S$ is the
Brunt-V\"{a}is\"{a}l\"{a} frequency, $S$ is the mean entropy, ${\bf
g}$ is the acceleration of gravity, ${\bf u}'$, $\, {\bf b}$ and
$s'$ are fluctuations of velocity, magnetic field and entropy (for
simplicity of notations we omitted prime in ${\bf b}$ because we did
not use new variables for magnetic fluctuations), ${\bf v}^{N}$, $\,
{\bf b}^{N} $ and $\, s^{N}$ are the nonlinear terms which include
the molecular viscous and diffusion terms, $p = p' + \sqrt{\rho_0}
\, ({\bf H} {\bf \cdot} {\bf b})$ are the fluctuations of total
pressure, $p'$ are the fluctuations of fluid pressure. Equations
(\ref{B1})-(\ref{B3}) for fluctuations of fluid velocity, entropy
and magnetic field are written in the anelastic approximation, which
is a combination of the Boussinesq approximation and the condition
${\bf\nabla}{\bf \cdot} \, (\rho_0 \, {\bf u}') = 0$. The equation,
${\bf\nabla}{\bf \cdot} \, {\bf u}' = \Lambda_\rho ({\bf u}' {\bf
\cdot} {\bf e})$, in the new variables reads: ${\bf\nabla}{\bf
\cdot} \, {\bf v} = (\Lambda_\rho / 2) ({\bf v} {\bf \cdot} {\bf
e})$, where $\bec{\nabla} \rho_0 / \rho_0 = - \Lambda_\rho {\bf e}$.
The quantities with the subscript $ "0" $ correspond to the
hydrostatic nearly isentropic basic reference state, i.e.,
$\bec{\nabla} P_{0} = \rho_{0} \, {\bf g} $ and $ {\bf g} {\bf
\cdot} [(\gamma P_{0})^{-1} \, \bec{\nabla} P_{0} - \rho_0^{-1}
\bec{\nabla} \rho_0] \approx 0$, where $\gamma$ is the specific
heats ratio and $P_{0}$ is the fluid pressure in the basic reference
state. The turbulent convection is regarded as a small deviation
from a well-mixed adiabatic reference state.

Using equations~(\ref{B1})-(\ref{B3}) written in a Fourier space we
derive equations for the two-point second-order correlation
functions of the velocity fluctuations $\langle v_i \, v_j\rangle$,
the magnetic fluctuations $\langle b_i \, b_j \rangle$, the entropy
fluctuations $\langle s \, s\rangle$, the cross-helicity $\langle
b_i \, v_j \rangle$, the turbulent heat flux $\langle s \, v_i
\rangle$ and $\langle s \, b_i \rangle$. The equations for these
correlation functions are given by equations (\ref{B6})-(\ref{B11})
in Appendix A. We split the tensor of magnetic fluctuations into
nonhelical, $h_{ij} = \langle b_i \, b_j \rangle$, and helical,
$h_{ij}^{(H)},$ parts. The helical part $h_{ij}^{(H)}$ depends on
the magnetic helicity and is determined by a dynamic equation which
follows from the magnetic helicity conservation arguments (see
below). We also split all second-order correlation functions,
$M^{(II)}$, into symmetric and antisymmetric parts with respect to
the wave vector ${\bf k}$, e.g., $h_{ij} = h_{ij}^{(s)} +
h_{ij}^{(a)}$, where the tensors $h_{ij}^{(s)} = [h_{ij}({\bf k}) +
h_{ij}(-{\bf k})] / 2$ describes the symmetric part of the tensor
and $h_{ij}^{(a)} = [h_{ij}({\bf k}) - h_{ij}(-{\bf k})] / 2$
determines the antisymmetric part of the tensor.

The second-moment equations include the first-order spatial
differential operators $\hat{\cal N}$  applied to the third-order
moments $M^{(III)}$. A problem arises how to close the system, i.e.,
how to express the set of the third-order terms $\hat{\cal N}
M^{(III)}$ through the lower moments $M^{(II)}$ (see, e.g., Orszag
1970, Monin and Yaglom 1975, McComb 1990). We use the spectral
$\tau$ approximation which postulates that the deviations of the
third-moment terms, $\hat{\cal N} M^{(III)}({\bf k})$, from the
contributions to these terms afforded by the background turbulent
convection, $\hat{\cal N} M^{(III,0)}({\bf k})$, are expressed
through the similar deviations of the second moments, $M^{(II)}({\bf
k}) - M^{(II,0)}({\bf k})$:
\begin{eqnarray}
\hat{\cal N} M^{(III)}({\bf k}) - \hat{\cal N} M^{(III,0)}({\bf k})
&=& - {1 \over \tau(k)} \, [M^{(II)}({\bf k})
\nonumber\\
&&- M^{(II,0)}({\bf k})] \;, \label{AAC3}
\end{eqnarray}
(see, e.g., Orszag 1970, Pouquet {\it et al.} 1976, Kleeorin {\it et
al.} 1990, Kleeorin {\it et al.} 1996, Blackman and Field 2002,
Kleeorin and Rogachevskii 2003, Rogachevskii and Kleeorin 2004,
Brandenburg {\it et al.} 2004, Brandenburg and Subramanian 2005b,
Kleeorin and Rogachevskii 2006), where $\tau(k)$ is the
scale-dependent relaxation time, which can be identified with the
correlation time of the turbulent velocity field. In the background
turbulent convection, the mean magnetic field is zero.

We apply the spectral $\tau$ approximation only for the nonhelical
part $h_{ij}$ of the tensor of magnetic fluctuations. The helical
part $h_{ij}^{(H)}$ depends on the magnetic helicity, and it is
determined by the dynamic equation which follows from the magnetic
helicity conservation arguments (see, e.g., Kleeorin and Ruzmaikin
1982, Gruzinov and Diamond 1994, Kleeorin {\it et al.} 1995,
Gruzinov and Diamond 1996, Kleeorin and Rogachevskii 1999, Kleeorin
{\it et al.} 2000, Blackman and Field 2000, Kleeorin {\it et al.}
2002, Blackman and Brandenburg 2002, Kleeorin {\it et al.} 2003,
Brandenburg and Subramanian 2005a, Zhang {\it et al.} 2006, and
references therein). The characteristic time of evolution of the
nonhelical part of the tensor $h_{ij}$ is of the order of the
turbulent time $\tau_{0} = l_{0} / u_{0}$, while the relaxation time
of the helical part of the tensor $h_{ij}^{(H)}$ is of the order of
$\tau_{0} \, {\rm Rm}$, where ${\rm Rm}= l_0 u_{0} / \eta$ is the
magnetic Reynolds number (which is very large), $u_{0}$ is the
characteristic turbulent velocity in the maximum scale of turbulent
motions $l_{0}$ and $\eta$ is the magnetic diffusivity due to
electrical conductivity of the fluid. In this study we consider an
intermediate nonlinearity which implies that the mean magnetic field
is not enough strong in order to affect the correlation time of
turbulent velocity field. The theory for a very strong mean magnetic
field can be corrected after taking into account a dependence of the
correlation time of the turbulent velocity field on the mean
magnetic field.

We assume also that the characteristic time of variation of the mean
magnetic field ${\bf B}$ is substantially larger than the
correlation time $\tau(k)$ for all turbulence scales. This allows us
to get a stationary solution for the equations for the second-order
moments, $M^{(II)}$. For the integration in ${\bf k}$ space of the
second-order moments we have to specify a model for the background
turbulent convection which is determined by equations
(\ref{K1})-(\ref{K2}) in Appendix A. This model takes into account
the inhomogeneity of the turbulence described by the two parameters:
$\Lambda_i^{(u)} = \nabla_i \langle {\bf u}'^2 \rangle^{(0)} /
\langle {\bf u}'^2 \rangle^{(0)}$ and $\Lambda_i^{(b)} = \nabla_i
\langle {\bf b}^2 \rangle^{(0)} / \langle {\bf b}^2 \rangle^{(0)}$.
This model includes also the inhomogeneity of the turbulent heat
flux, $\Lambda_i^{(F)} = \nabla_i \langle |{\bf u}'| \, s'
\rangle^{(0)} / \langle |{\bf u}'| \,  s' \rangle^{(0)}$, and the
inhomogeneity of the fluid density described by the parameter
$\Lambda_\rho$. The quantities with the superscript $(0)$ correspond
to the background turbulent convection with ${\bf B} =0$. Using the
solution of the derived second-moment equations, we determine the
nonlinear electromotive force, ${\cal E}_{i} = \varepsilon_{imn} \,
\int \langle b_n \, v_m \rangle_{\bf k} \,{\rm d} {\bf k}$, in the
turbulent convection (see Appendix A), where $\varepsilon_{ijk}$ is
the fully antisymmetric Levi-Civita tensor. This allowed us to
determine the nonlinear drift velocities of the mean magnetic field
and nonlinear turbulent magnetic diffusion, and to study the
small-scale magnetic buoyancy and magnetic pumping effects in the
turbulent convection.

\section*{3. The axisymmetric dynamo}

Let us consider the axisymmetric $\alpha \Omega$ dynamo problem. The
mean magnetic field in the local system of coordinate is ${\bf B} =
B(x,z){\bf e}_y + \bec{\nabla} {\bf \times} [A(x,z){\bf e}_y]$,
where $B(x,z)$ and $A(x,z)$ are determined by the dimensionless
equations
\begin{eqnarray}
{\partial A \over \partial t} &=& \alpha({\bf B}) B +
\eta_{_{A}}^{(z)}({\bf B}) {\partial^2 A \over \partial z^2} +
\eta_{_{A}}^{(x)}({\bf B}) {\partial^2 A \over \partial x^2}
\nonumber \\
&& - ({\bf V}_{A}({\bf B}) {\bf \cdot} \bec{\nabla}) A \;,
\label{C1} \\
{\partial B \over \partial t} &=& D \, [\bec{\nabla}(\delta \Omega)
{\bf \times} \bec{\nabla} A]_y + \bec{\nabla} {\bf \cdot} [\bec{\hat
\eta}_{_{B}}({\bf B}) \bec{\nabla} B
\nonumber \\
&& - {\bf V}_{B}({\bf B}) B] \;, \label{C2}
\end{eqnarray}
$\delta \Omega$ determine the differential rotation, $D$ is the
dynamo number (see below), $\alpha({\bf B})$ is the total
(hydrodynamic + magnetic) nonlinear alpha effect (see, e.g.,
Kleeorin {\it et al.} 2000, Rogachevskii and Kleeorin 2000, and
references therein), $\bec{\hat \eta}_{_{B}}$ is the diagonal tensor
with the components $\eta_{_{B}}^{(z,x)}({\bf B})$ of the nonlinear
turbulent magnetic diffusion of toroidal field,
$\eta_{_{A}}^{(z,x)}({\bf B})$ are the nonlinear turbulent magnetic
diffusion coefficients of the poloidal magnetic field, ${\bf
V}_{A}({\bf B})$ and ${\bf V}_{B}({\bf B})$ are the nonlinear drift
velocities (see below). The axis $z$ of the local system of
coordinate is directed opposite to the gravity acceleration ${\bf
g}$ and the axis $x$ is in meridional plane and directed to the
equator, so that the spherical coordinates $(r, \theta, \varphi)$
translate to the local system of coordinate $(z, x, y)$.

We adopt here the dimensionless form of the mean dynamo equations;
in particular, length is measured in units of $L$, time is measured
in units of the turbulent magnetic diffusion time $L^{2} /
\eta_{_{T}}$ and ${\bf B}$ is measured in units of the equipartition
energy $B_{\rm eq} = \sqrt{\mu \rho_0} \, u_0$, $\, \alpha$  is
measured  in units of $\alpha_\ast$ (the maximum value of the
hydrodynamic part of the $\alpha$ effect), the nonlinear turbulent
magnetic diffusion coefficients are measured in units of
$\eta_{_{T}}  = l_0 u_{0} / 3$, the nonlinear drift velocities ${\bf
V}_{A,B}({B})$ are measured in the units of $\eta_{_{T}} / L$, the
differential rotation $\delta \Omega$ is measured in units of
$\delta \Omega_\ast$ and the dimensionless parameters
$\Lambda^{(u)}$, $\, \Lambda^{(b)}$, $\, \Lambda_\rho$ and
$\Lambda^{(F)}$ are measured in the units of $L^{-1}$. We define
$R_\alpha = L \alpha_\ast / \eta_{_{T}} ,$ $\, R_\omega = r \, ({\rm
d} (\delta\Omega_\ast)/{\rm d}r) \, L^2/\eta_{_{T}}$, and the dynamo
number $D = R_\omega R_\alpha$.

The derivation of equation for the nonlinear electromotive force
allows us to determine the nonlinear turbulent magnetic diffusion
coefficients and the nonlinear drift velocities of the mean magnetic
field, which are given by
\begin{eqnarray}
\eta_{_{A,B}}^{(z)}({\bf B}) &=& \eta_{_{A,B}}^{(v)}({\bf B}) +
a_\ast \eta_{_{A,B}}^{(F,z)}({\bf B}) \;,
\nonumber\\
\eta_{_{A,B}}^{(x)}({\bf B}) &=& \eta_{_{A,B}}^{(v)}({\bf B}) +
a_\ast \eta_{_{A,B}}^{(F,x)}({\bf B}) \;,
\nonumber\\
{\bf V}_{A,B}({\bf B}) &=& {\bf V}_{A,B}^{(v)}({\bf B}) + a_\ast
{\bf V}_{A,B}^{(F)}({\bf B}) \; .
\label{C9}
\end{eqnarray}
Here the superscript $(v)$ corresponds to the contributions from the
purely hydrodynamic turbulence and the superscript $(F)$ corresponds
to the contributions from the turbulent heat flux. These
contributions are given by equations~(\ref{AA1})-(\ref{AA14}) in
Appendix A. The parameter $a_\ast$ which is determined by the budget
equation for the total energy, is given by
\begin{eqnarray}
a_\ast^{-1} = 1 + { \nu_{_{T}} (\nabla \, U)^2 + \eta_{_{T}} (\nabla
\, B)^2 /(\rho \mu) \over g \, F_\ast} \;,
 \label{AAC1}
\end{eqnarray}
where ${\bf U}$ is the mean velocity and $\nu_{_{T}}$ is the
turbulent viscosity.

\begin{figure}
\centering
\includegraphics[width=8cm]{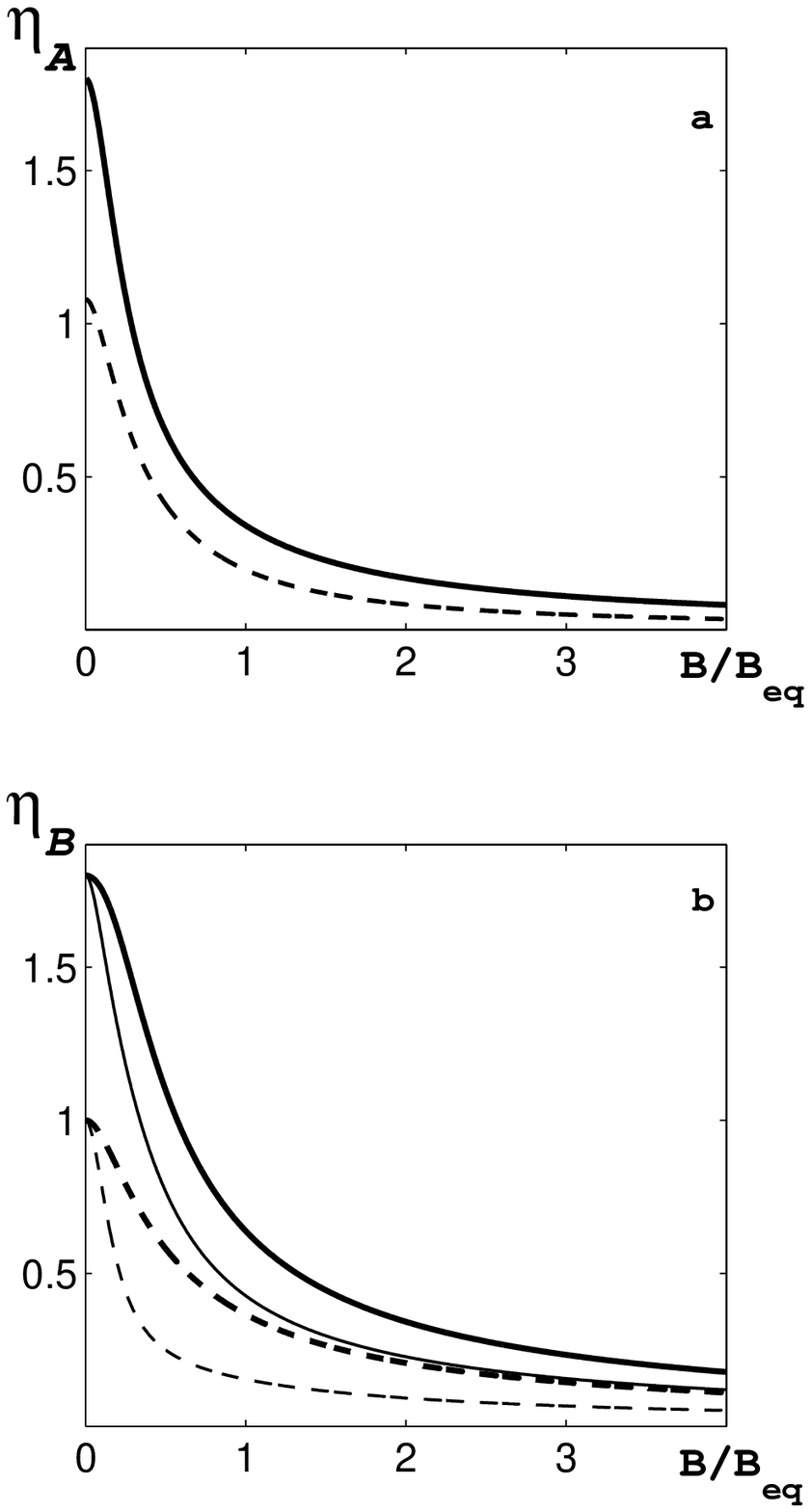}
\caption{\label{Fig1} Nonlinear turbulent magnetic diffusion
coefficients (a) $\eta_{_{A}}$ and (b) $\eta_{_{B}}$ in the vertical
(solid) and horizontal (dashed) directions in a turbulent convection
with $a_\ast = 0.8$. The thin curves in figure~1b correspond to
$\epsilon=0$ and thick curves to $\epsilon=1$.}
\end{figure}

\begin{figure}
\centering
\includegraphics[width=8cm]{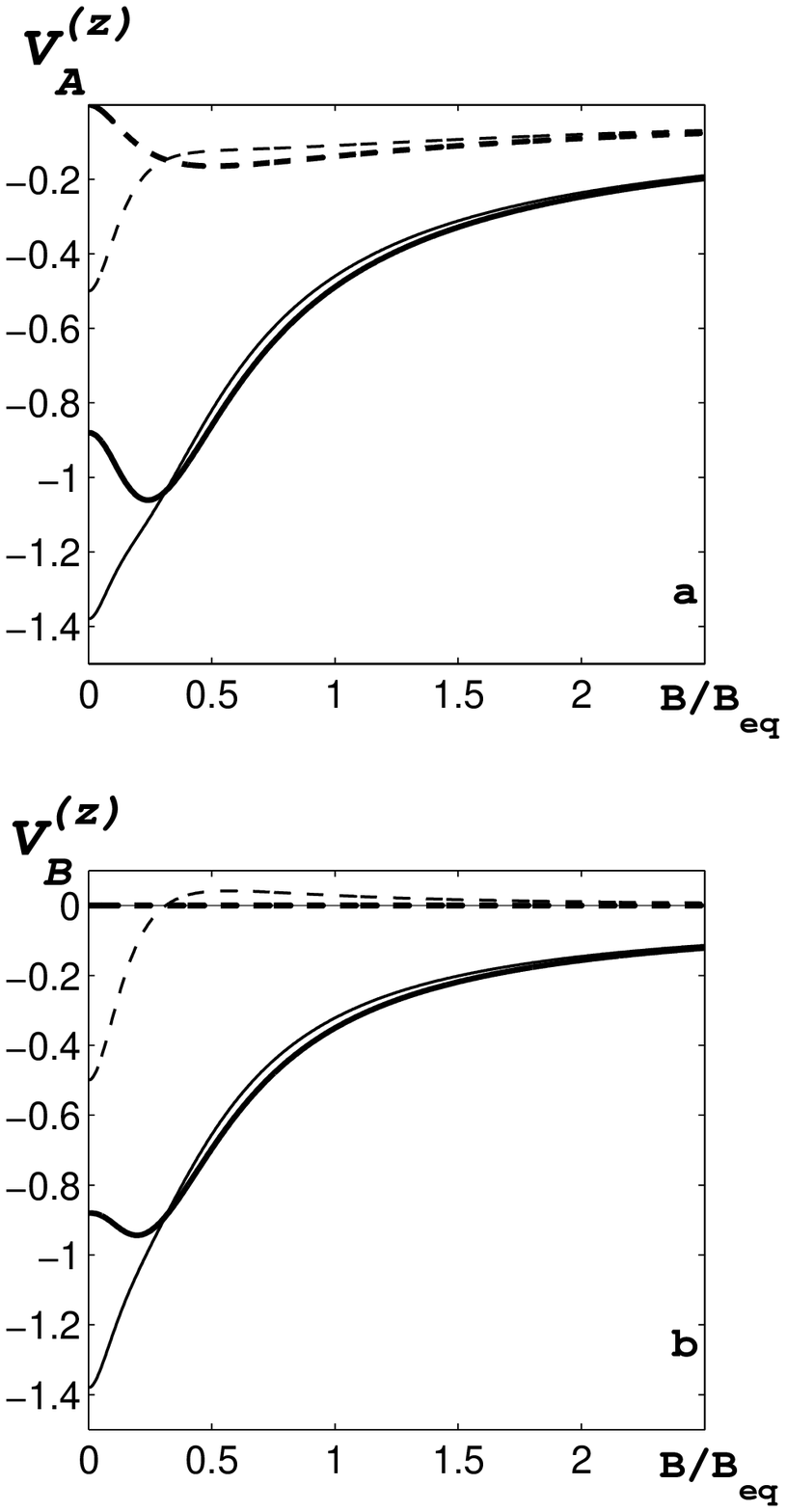}
\caption{\label{Fig2} Vertical nonlinear drift velocities (a)
$V_A^{(z)}$ and (b) $V_B^{(z)}$ in a turbulent convection with
$a_\ast = 0.8$ (solid) and in a nonconvective turbulence, $a_\ast =
0$ (dashed) for $\Lambda^{(u)}_z = \Lambda^{(F)}_z = \Lambda^{(B)}_z
= \Lambda_\rho = \Lambda^{(b)}_z =1$.  The thin curves correspond to
$\epsilon=0$ and thick curves to $\epsilon=1$.}
\end{figure}

The asymptotic formulae for the nonlinear turbulent magnetic
diffusion coefficients and the nonlinear drift velocities for the
weak mean magnetic fields, $B \ll B_{\rm eq} / 4$, are given by
\begin{eqnarray}
\eta_{_{A}}^{(z)}({B}) &=& \eta_{_{B}}^{(z)}({B}) = 1 + a_\ast \;,
\nonumber\\
\eta_{_{A}}^{(x)}({B}) &=& 1 + 0.1 a_\ast \;, \; \;
\eta_{_{B}}^{(x)}({B}) = 1 \;,
\nonumber\\
V_{A}^{(z)}({B}) &=& V_{B}^{(z)}({B}) = - {1 \over 2}
\biggl[\Lambda_z^{(u)} - \epsilon \Lambda_z^{(b)}
\nonumber \\
&& - \epsilon \Lambda_\rho + {9 \, a_\ast \over 5} \Lambda_\rho
\biggr] \;,
\nonumber\\
V_{A}^{(x)}({B}) &=& V_{B}^{(x)}({B}) = - {1 \over 2}
[\Lambda_x^{(u)} - \epsilon \Lambda_x^{(b)}] \;, \label{AAC9}
\end{eqnarray}
where we neglect the terms $\sim {\rm O}(\beta^{2})$. Here $\beta =
\sqrt{8} \, B /B_{\rm eq}$ and the parameter $\epsilon = \langle
{\bf b}^2 \rangle^{(0)} / \langle {\bf v}^2 \rangle^{(0)} $. When $B
\gg B_{\rm eq} / 4$ the nonlinear turbulent magnetic diffusion
coefficients and the nonlinear drift velocities are given by
\begin{eqnarray}
\eta_{_{A}}^{(z)}({B}) &=& {a_\ast \over \beta}  \;, \quad
\eta_{_{A}}^{(x)}({B}) = {2 \, a_\ast \over 5 \beta} \;,
\nonumber\\
\eta_{_{B}}^{(z)}({B}) &=& {2 \, (1 + \epsilon) \over 3 \beta} +
{a_\ast \over \beta} \;,
\nonumber\\
\eta_{_{B}}^{(x)}({B}) &=& {2 \, (1 + \epsilon) \over 3 \beta} \;,
\quad {\bf V}_{B}({B}) =  - {a_\ast \over \beta} \Lambda_\rho {\bf
e} \;,
\nonumber\\
{\bf V}_{A}({B}) &=& -  {1 + \epsilon \over 3 \beta} {\bf
\Lambda^{(B)}} - {a_\ast \over \beta} \Lambda_\rho {\bf e} \;,
\label{AAC10}
\end{eqnarray}
where we neglect the terms $\sim {\rm O}(\beta^{-2})$.

The nonlinear turbulent magnetic diffusion coefficients
$\eta_{_{A,B}}^{(z,x)}$ of the poloidal and toroidal components of
the mean magnetic field in the vertical (along the $z$-axis) and
horizontal (along the $x$-axis) directions are shown in figure~1 for
the turbulent convection with $a_\ast = 0.8$. The magnetic
fluctuations due to the small-scale dynamo (described by the
parameter $\epsilon$) increase the turbulent magnetic diffusion of
the toroidal mean magnetic field (see figure~1b), and they do not
affect the turbulent magnetic diffusion of the poloidal field. Note
also that the nonlinear turbulent magnetic diffusion in a turbulent
convection is anisotropic even for a weak mean magnetic field. In
particular, it is enhanced in the vertical (radial) direction.

The vertical nonlinear drift velocities of poloidal and toroidal
components of the mean magnetic field in the turbulent convection
$(a_\ast = 0.8)$ and in the nonconvective turbulence $(a_\ast = 0)$
are shown in figure~2. The turbulent convection enhances the
nonlinear drift velocities of the mean magnetic field in comparison
with the case of a purely hydrodynamic turbulence (see figure~2). In
the next section we discuss the nonlinear drift velocities of the
mean magnetic field in the solar convective zone which cause the
small-scale magnetic buoyancy and magnetic pumping effects.

\section*{4. Discussion}

\begin{figure}
\centering
\includegraphics[width=8cm]{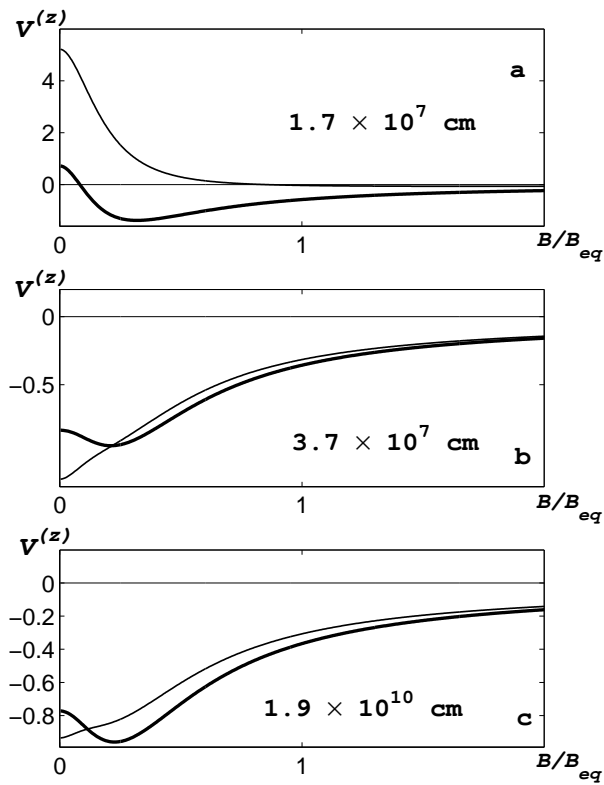}
\caption{\label{Fig3} Vertical nonlinear drift velocities
$(V_B^{(z)}= V_A^{(z)} \equiv V^{(z)})$ in a turbulent convection
with $a_\ast = 0.8$ for $\Lambda^{(b)}_z = \Lambda^{(B)}_z =0$, and
for different depths $h$ of the convective zone (from the solar
surface): (a) $h=1.7 \times 10^7$ cm; (b) $h=3.7 \times 10^7$ cm;
(c) $h=1.9 \times 10^{10}$ cm. The thin curves correspond to
$\epsilon=0$ and thick curves to $\epsilon=1$.}
\end{figure}

\begin{figure}
\centering
\includegraphics[width=8cm]{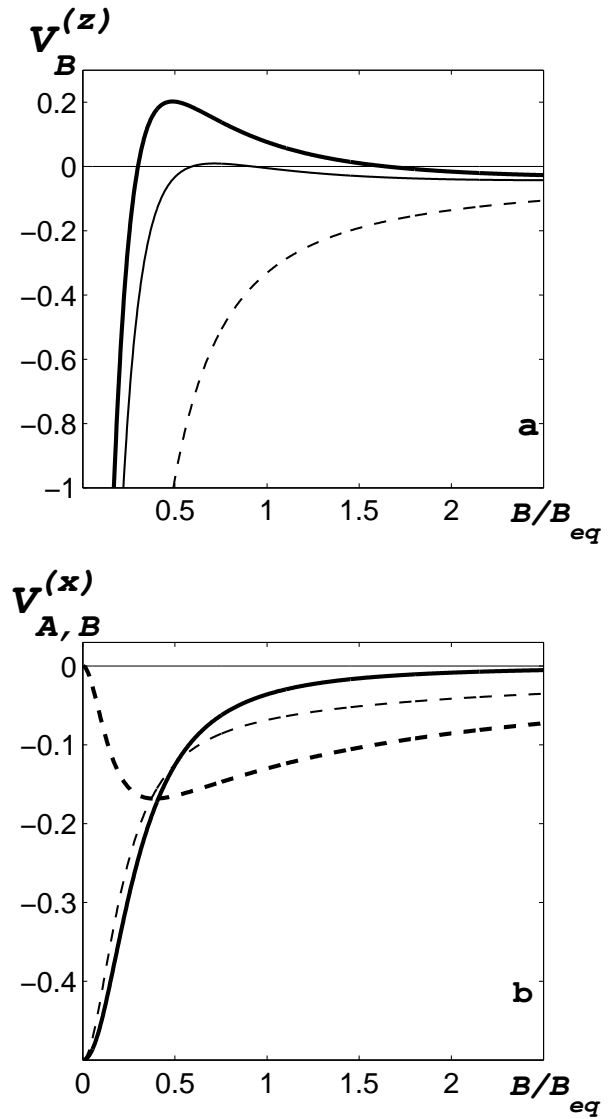}
\caption{\label{Fig4} Vertical nonlinear drift velocity $V_B^{(z)}$
(figure~4a) of toroidal magnetic field in the overshoot layer with
$a_\ast = 0.8$ for $\Lambda^{(u)}_z = \Lambda^{(F)}_z = 20$, $\,
\Lambda_\rho =\Lambda^{(B)}_z = 1$, $\, \Lambda^{(b)}_z =
\Lambda^{(u)}_z -\Lambda_\rho$. The thick curve corresponds to
$\epsilon=1$, the thin solid curve corresponds to $\epsilon=0.9$ and
thin dashed curve corresponds to $\epsilon=0.5$. Horizontal
nonlinear effective drift velocities $V_{A,B}^{(x)}$ (figure~4b) of
toroidal (solid) and poloidal (dashed) magnetic fields in the
turbulent convection with $a_\ast = 0.8$ for $\Lambda^{(u)}_x
=\Lambda^{(b)}_x = \Lambda^{(F)}_x = \Lambda^{(B)}_x = 1$. The thin
curves correspond to $\epsilon=0$ and thick curves to $\epsilon=1$.}
\end{figure}

Let us discuss the small-scale magnetic buoyancy and magnetic
pumping effects. In figure~3 the vertical nonlinear drift velocities
of the toroidal and poloidal magnetic fields are plotted for
different depths $h$ of the solar convective zone (measured from the
solar surface): $h=1.7 \times 10^7$ cm (figure~3a); $h=3.7 \times
10^7$ cm (figure~3b), and $h=1.9 \times 10^{10}$ cm (figure~3c). In
order to estimate the governing parameters we use the models of the
solar convective zone (see, e.g., Spruit 1974, Baker and Temesvary
1966). More modern treatments make little difference to these
estimates.

In particular, in the upper part of the solar convective zone, say
at the depth $ h_\ast \sim 1.7 \times 10^7$ cm, the parameters are
as follows: the characteristic turbulent velocity $u_0 \sim 2.2
\times 10^5 $ cm s$^{-1}$; the maximum scale of turbulent motions $
l_0 \sim 3.3 \times 10^7$ cm; the fluid density $\rho \sim 4.6
\times 10^{-7}$ g cm$^{-3}$; the turbulent magnetic diffusion
$\eta_{_{T}} \, \sim 2.4 \times 10^{12} $ cm$^2$ s$^{-1}$; the
density stratification scale $\Lambda_\rho^{-1} \sim 10^8$ cm and
the characteristic scale of the inhomogeneity of the turbulent
magnetic diffusion $\Lambda_\eta^{-1} = |\nabla_r \, \eta_{_{T}} /
\eta_{_{T}}|^{-1} \sim 10^7$ cm.

At the depth $ h_\ast \sim 3.7 \times 10^7$ cm, the parameters are
$u_0 \sim 1.5 \times 10^5 $ cm s$^{-1}$; $\, l_0 \sim 4.5 \times
10^7$ cm; $\, \rho \sim 8.3 \times 10^{-7}$ g cm$^{-3}$; $\,
\eta_{_{T}} \, \sim 2.3 \times 10^{12} $ cm$^2$ s$^{-1}$; $\,
\Lambda_\rho^{-1} \sim 4 \times 10^7$ cm and $\, \Lambda_\eta^{-1}
\sim 2.2 \times 10^8$ cm.

At the bottom of the solar convective zone, say at the depth $
h_\ast \sim 1.9 \times 10^{10}$ cm, the parameters are $u_0 \sim 2
\times 10^3 $ cm s$^{-1}$; $\, l_0 \sim 8.1 \times 10^9$ cm; $\,
\rho \sim 2.1 \times 10^{-1}$ g cm$^{-3}$; $\, \eta_{_{T}} \, \sim
5.2 \times 10^{12} $ cm$^2$ s$^{-1}$; $\, \Lambda_\rho^{-1} \sim 6.5
\times 10^9$ cm and $\, \Lambda_\eta^{-1} \sim 8 \times 10^{10}$ cm.

Figure~3 demonstrates that only near the solar surface the radial
drift velocity for a weak mean magnetic field is directed upward to
the surface of the Sun. This causes to the small-scale magnetic
buoyancy effect. However, in the main part of the solar convective
zone the radial nonlinear drift velocities of the toroidal and
poloidal mean magnetic fields are directed downward. This results in
the small-scale magnetic pumping effect. These phenomena are
determined by the nonlinear drift velocities in the nonlinear
electromotive force, and they are different from the large-scale
magnetic buoyancy and magnetic pumping effects. The large-scale
phenomena are caused by the effect of the mean magnetic field on the
large-scale density stratified fluid flow. These large-scale
phenomena are stronger than the small-scale magnetic buoyancy and
magnetic pumping effects when the mean magnetic field is larger than
the equipartition field. In particular, the ratio of the velocities
which correspond to the large-scale and small-scale effects, is of
the order of $(B / B_{\rm eq})^2$.

In figure~4 the vertical (figure~4a) and horizontal (figure~4b)
nonlinear drift velocities of the toroidal mean magnetic field are
plotted for the overshoot layer located at the bottom of the solar
convective zone. In these layer the turbulence and the turbulent
heat flux are strongly inhomogeneous. The drift velocities in
figures~2-4 are measured in the units of $ \eta_{_{T}} \,
\Lambda_\rho $. Here we assumed that $\Lambda^{(b)}_z =
\Lambda^{(u)}_z -\Lambda_\rho$, which implies that $\Lambda_i^{(b)}
= \nabla_i (\rho_0 \, \langle {\bf u}'^2 \rangle^{(0)}) / (\rho_0 \,
\langle {\bf u}'^2 \rangle^{(0)})$. Figure~4a demonstrates that the
vertical nonlinear drift velocity of the toroidal mean magnetic
field depends strongly on the level of the magnetic fluctuations
caused by the small-scale dynamo (described by the parameter
$\epsilon$). If there is a small deviation from $\epsilon=1$ (the
equipartition between the kinetic and magnetic turbulent energies)
there is only the magnetic pumping effect in the overshoot layer. On
the other hand, the horizontal nonlinear drift velocity of the
toroidal mean magnetic field in the overshoot layer is negative,
i.e., it is directed to the solar polar regions (see figure~4b).

The magnetic pumping in three-dimensional compressible rotating
magnetoconvection has been studied by Ossendrijver {\it et al.}
(2002) in direct numerical simulations (see also review by
Ossendrijver 2003). The resulting pumping effects are isolated by
calculating of the effective drift velocities in turbulent
convection. The pumping effects act differently on different
components of the mean magnetic field (Ossendrijver {\it et al.}
2002). This result is in a good agreement with our results [see
figure~2 and equations~(\ref{C1}), (\ref{C2}), (\ref{AAC10}),
(\ref{AA3}), (\ref{AA4})]. The pumping effect in the vertical
direction is found to be equivalent to a predominating downward
advection with a maximum drift velocity of the order of 10\% of the
turbulent velocity (Ossendrijver {\it et al.} 2002). This is in
agreement with our theoretical findings (see, e.g., figures~2, 3 and
4a). Note that the effective drift velocity due to the inhomogeneity
of the fluid density (see Kichatinov and R\"{u}diger 1992) also
causes a predominating downward drift of the mean magnetic field.

The small-scale magnetic pumping and buoyancy effects were
investigated in the present study for large hydrodynamic and
magnetic Reynolds numbers using the spectral $\tau$ approximation
(the third-order closure procedure). Previous analytical studies of
the small-scale magnetic pumping and buoyancy effects (see
Kichatinov 1991, Kichatinov and R\"{u}diger 1992, Kichatinov and
Pipin 1993) were performed using the second order correlation
approximation (SOCA). This approximation is valid for small
hydrodynamic Reynolds numbers. Indeed, even in a highly conductivity
limit (large magnetic Reynolds numbers) SOCA is valid only for small
Strouhal numbers, while for large hydrodynamic Reynolds numbers
(fully developed turbulence) the Strouhal number is 1. In the
present study we take into account the inhomogeneity of the fluid
density assuming that $\langle \rho u'_i u'_j \rangle$ is weakly
inhomogeneous (see equation~(\ref{K1})). This is in agreement with
the models of the solar convective zone (see, e.g., Spruit 1974,
Baker and Temesvary 1966). On the other hand, in studies by
Kichatinov (1991) and Kichatinov and R\"{u}diger (1992) it was
assumed that $\langle \rho^2 u'_i u'_j \rangle$ is weakly
inhomogeneous. Since the density in the solar convective zone varies
over six orders of magnitude, the validity of the latter suggestion
is questionable.

\section*{5. Conclusions}

In summary, we study the nonlinear drift of the mean magnetic field
in a turbulent convection. Three kinds of the inhomogeneities
determine the nonlinear drift velocities of the mean magnetic field:
(i) the inhomogeneous turbulence; (ii) the nonuniform fluid density
and (iii) the nonuniform turbulent heat flux. The inhomogeneous
turbulence causes the well-known turbulent diamagnetic and
paramagnetic velocities. The nonlinear drift velocities of the mean
magnetic field result in the small-scale magnetic buoyancy and
magnetic pumping effects in the turbulent convection. In the main
part of the solar convective zone the small-scale magnetic pumping
effect dominates (i.e., the radial nonlinear drift velocity of the
mean magnetic field is directed downward to the bottom of the
convective zone), while near the solar surface the small-scale
magnetic buoyancy effect is important when the mean magnetic field
is weak. These small-scale phenomena can be stronger than the
large-scale magnetic pumping and magnetic buoyancy which are caused
by the influence of the mean magnetic field on the stratified fluid
flow.

\bigskip
\bigskip
{\bf Acknowledgment}

An important part of this work was performed during our visit to the
Isaac Newton Institute for Mathematical Sciences (University of
Cambridge) during the programme "Magnetohydrodynamics of Stellar
Interiors".

\bigskip
\leftline{\bf References}
\bigskip
{
\parindent 0em
  \hangindent 2em
  \parskip2ex

    Baker N. and Temesvary, S., {\it Tables of Convective Stellar
Envelope Models}, 1966 (NASA: New York).

    Blackman, E.G. and Brandenburg, A., Dynamic nonlinearity in large
scale dynamos with shear. {\it Astrophys. J.}, 2002, {\bf 579},
359-373.

    Blackman, E.G. and Field, G., Constraints on the magnitude of alpha
in dynamo theory. {\it Astrophys. J.}, 2000, {\bf 534}, 984-988.

    Blackman, E.G. and Field, G., New dynamical mean-field dynamo
theory and closure approach. {\it Phys. Rev. Lett.}, 2002, {\bf 89},
265007 (1-4).

    Brandenburg, A., The case for a distributed solar dynamo
shaped by near-surface shear. {\it Astrophys. J.}, 2005, {\bf 625},
539-547.

    Brandenburg, A., K\"{a}pyl\"{a}, P. and Mohammed, A., Non-Fickian
diffusion and tau-approximation from numerical turbulence. {\it
Phys. Fluids}, 2004, {\bf 16}, 1020-1027.

    Brandenburg, A. and Subramanian, K., Astrophysical magnetic fields
and nonlinear dynamo theory. {\it Phys. Rept.}, 2005a,  {\bf 417},
1-209.

    Brandenburg, A. and Subramanian, K., Minimal
tau approximation and simulations of the alpha effect. {\it Astron.
Astrophys.}, 2005b, {\bf 439}, 835-843.

    Drobyshevski, E.M. and Yuferev, V.S., Topological pumping of
magnetic flux by three-dimensional convection. {\it J. Fluid Mech.},
1974, {\bf 65}, 33-44.

    Galloway, D.J. and Proctor, M.R.E., The kinematics of hexagonal
magnetoconvection. {\it Geophys. Astrophys. Fluid Dyn.},  1983, {\bf
24}, 109-136.

    Gilman, P.A., Instability of magnetohydrostatic stellar interiors
from magnetic buoyancy. {\it Astrophys. J.}, 1970, {\bf 162},
1019-1029.

    Gruzinov, A.  and Diamond, P. H., Self-consistent theory of
mean-field electrodynamics. {\it Phys. Rev. Lett.}, 1994, {\bf 72},
1651-1653.

    Gruzinov, A.  and Diamond, P.H., Nonlinear mean-field
electrodynamics of turbulent dynamos. {\it Phys. Plasmas}, 1996,
{\bf 3}, 1853-1857.

    Kichatinov, L. L., Turbulent transport of magnetic fields
in a highly conducting rotating fluid and the solar cycle. {\it
Astron. Astrophys.}, 1991, {\bf 243}, 483-491.

    Kichatinov, L.L. and Pipin, V.V., Mean-field buoyancy.
{\it Astron. Astrophys.}, 1993, {\bf 274}, 647-652.

    Kichatinov, L.L. and R\"{u}diger, G., Magnetic-field advection
in inhomogeneous turbulence. {\it Astron. Astrophys.}, 1992, {\bf
260}, 494-498.

    Kleeorin N., Kuzanyan, K., Moss D., Rogachevskii I., Sokoloff D. and
Zhang, H., Magnetic helicity evolution during the solar activity
cycle: observations and dynamo theory. {\it Astron. Astrophys.},
2003, {\bf 409}, 1097-1105.

    Kleeorin, N., Mond, M. and Rogachevskii, I., Magnetohydrodynamic
turbulence in the solar convective zone as a source of oscillations
and sunspots formation. {\it Astron. Astrophys.}, 1996, {\bf 307},
293-309.

    Kleeorin N., Moss D., Rogachevskii I. and Sokoloff D., Helicity
balance and steady-state strength of dynamo generated galactic
magnetic field. {\it Astron. Astrophys.}, 2000, {\bf 361}, L5-L8.

    Kleeorin N., Moss D., Rogachevskii I. and Sokoloff D., The role of
magnetic helicity transport in nonlinear galactic dynamos. {\it
Astron. Astrophys.}, 2002, {\bf 387}, 453-462.

    Kleeorin, N. and Rogachevskii, I., Magnetic helicity tensor for an
anisotropic turbulence. {\it Phys. Rev. E}, 1999, {\bf 59},
6724-6729.

    Kleeorin, N. and Rogachevskii, I., Effect of rotation on a
developed turbulent stratified convection: the hydrodynamic
helicity, the alpha-effect and the effective drift velocity. {\it
Phys. Rev. E}, 2003, {\bf 67}, 026321 (1-19).

    Kleeorin, N. and Rogachevskii, I., Effect of heat flux on
differential rotation in turbulent convection. {\it Phys. Rev. E},
2006, {\bf 73}, 046303 (1-9).

    Kleeorin, N., Rogachevskii, I. and Ruzmaikin, A., Magnetic force
reversal and instability in a plasma with advanced
magnetohydrodynamic turbulence. {\it Sov. Phys. JETP}, 1990, {\bf
70}, 878-883. Translation from {\it Zh. Eksp. Teor. Fiz.}, 1990,
{\bf 97}, 1555-1565.

    Kleeorin, N., Rogachevskii, I. and Ruzmaikin, A., Magnitude of
dynamo - generated magnetic field in solar - type convective zones.
{\it Astron. Astrophys.}, 1995, {\bf 297}, 159-167.

    Kleeorin, N. and Ruzmaikin, A., Dynamics of the averaged turbulent
helicity in a magnetic field. {\it Magnetohydrodynamics}, 1982, {\bf
18}, 116-122. Translation from {\it Magnitnaya Gidrodinamika}, 1982,
{\bf 2}, 17-24.

    Krause, F. and R\"{a}dler, K.-H., {\it Mean-Field
Magnetohydrodynamics and  Dynamo Theory}, 1980 (Pergamon: Oxford).

    Kulsrud, R., A critical review of galactic dynamos. {\it Annu.
Rev. Astron. Astrophys.}, 1999, {\bf 37}, 37-64.

    McComb, W.D., {\it The Physics of Fluid Turbulence}, 1990
(Clarendon: Oxford).

    Moffatt, H.K., {\it Magnetic Field Generation in Electrically
Conducting  Fluids}, 1978 (Cambridge University Press: New York).

    Monin, A.S. and Yaglom, A.M., {\it Statistical Fluid Mechanics}, 1975, Vol. 2
(MIT Press: Cambridge, Massachusetts).

    Moreno-Insertis, F., Sch\"{u}ssler, M. and Ferriz-Mas, A., Enhanced
inertia of thin magnetic flux tubes. {\it Astron. Astrophys.}, 1996,
{\bf 312}, 317-326.

    Orszag, S.A., Analytical theories of turbulence. {\it J. Fluid
Mech.}, 1970, {\bf 41}, 363-386.

    Ossendrijver, M., Stix, M., Brandenburg, A., R\"{u}diger, G.,
Magnetoconvection and dynamo coefficients. II. Field-direction
dependent pumping of magnetic field. {\it Astron. Astrophys.}, 2002,
{\bf 394}, 735-745.

    Ossendrijver, M., The solar dynamo. {\it Astron. Astrophys. Rev.},
2003, {\bf 11}, 287-367.

    Parker, E., The formation of sunspots from the solar toroidal
field. {\it Astrophys. J.}, 1955, {\bf 121}, 491-507.

    Parker, E., The dynamical state of the interstellar gas and
fluids. {\it Astrophys. J.}, 1966, {\bf 145}, 811-833.

    Parker, E., {\it Cosmical Magnetic Fields}, 1979 (Oxford University
Press: New York).

    Pouquet, A., Frisch, U. and Leorat, J.,  Strong MHD turbulence and
the nonlinear dynamo effect. {\it J. Fluid Mech.}, 1976, {\bf 77},
321-354.

    Priest, E.R., {\it Solar Magnetohydrodynamics}, 1982 (D. Reidel Publ.
Co.: Dordrecht).

    R\"{a}dler, K.-H., Kleeorin, N. and Rogachevskii, I., The mean
electromotive force for MHD turbulence: the case of a weak mean
magnetic field and slow rotation. {\it Geophys. Astrophys. Fluid
Dynam.}, 2003, {\bf 97}, 249-274.

    Roberts, P.H. and Soward, A.M., A unified approach to mean field
electrodynamics. {\it Astron. Nachr.}, 1975, {\bf 296}, 49-64.

    Roberts, P.H. and Soward, A.M., Dynamo theory. {\it Annu. Rev.
Fluid Mech.}, 1992, {\bf 24}, 459-512.

    Rogachevskii, I. and  Kleeorin, N., Electromotive force for an
anisotropic turbulence: intermediate nonlinearity {\it Phys. Rev.
E}, 2000, {\bf 61}, 5202-5210.

    Rogachevskii, I. and  Kleeorin, N., Nonlinear theory of a
"shear-current" effect and mean-field magnetic dynamos. {\it Phys.
Rev. E}, 2004, {\bf 70}, 046310 (1-15).

    Ruzmaikin, A.A., Shukurov, A.M. and Sokoloff, D.D., {\it Magnetic
Fields of Galaxies}, 1988 (Kluwer Academic: Dordrecht).

    Sch\"{u}ssler, M., Caligari, P., Ferriz-Mas, A. and
Moreno-Insertis, F., Instability and eruption of magnetic flux tubes
in the solar convection zone. {\it Astron. Astrophys.}, 1994, {\bf
281}, L69-L72.

    Spiegel, E.A. and Weiss, N.O., Magnetic activity and variations
in solar luminosity. {\it Nature}, 1980, {\bf 287}, 616-617.

    Spruit, H.C., A model of solar convective zone. {\it Solar
Phys.}, 1974, {\bf 34}, 277-290.

    Spruit, H. C., Motion of magnetic flux tubes in the solar
convection zone and chromosphere. {\it Astron. Astrophys.}, 1981,
{\bf 98}, 155-160.

    Spruit, H.C. and van Ballegooijen, A.A., Stability of toroidal
flux tubes in stars. {\it Astron. Astrophys.}, 1982, {\bf 106},
58-66.

    Stix, M., {\it The Sun: An Introduction}, 1989 (Springer:
Berlin and Heidelberg).

    Thomas, J.H., Weiss, N.O., Tobias, S.M. and Brummell, N.H.,
Downward pumping of magnetic flux as the cause of filamentary
structures in sunspot penumbrae. {\it Nature}, 2002, {\bf 420},
390-393.

    Tobias, S.M., Brummell, N.H., Clune T.L. and Toomre, J.,
Transport and storage of magnetic field by overshooting turbulent
compressible convection. {\it Astrophys. J.}, 2001, {\bf 549},
1183-1203.

    Tobias, S.M. and Hughes, D.W., The influence of velocity shear on
magnetic buoyancy instability in the solar tachocline. {\it
Astrophys. J.}, 2004, {\bf 603}, 785-802.

    Vainshtein, S.I. and Kichatinov, L.L., The macroscopic
magnetohydrodynamics of inhomogeneously turbulent cosmic plasmas.
{\it Geophys. Astrophys. Fluid Dynam.}, 1983, {\bf 24}, 273-298.

    Weiss, N.O., The expulsion of magnetic flux by eddies.
{\it Proc. R. Soc. Lond.}, 1966, {\bf A293}, 310-328.

    Zeldovich, Ya.B., The magnetic field in the two-dimensional
motion of a conducting turbulent fluid. {\it Sov. Phys. JETP}, 1957,
{\bf 4}, 460-462. Translation from {\it Zh. Eksp. Teor. Fiz.}, 1956,
{\bf 31}, 154-156.

    Zeldovich, Ya.B.,  Ruzmaikin, A. A. and Sokoloff, D. D., {\it
Magnetic Fields in Astrophysics}, 1983 (Gordon and Breach: New
York).

    Zhang, H., Sokoloff, D., Rogachevskii, I.,  Moss, D.,
Lamburt, V., Kuzanyan, K. and Kleeorin N., The Radial Distribution
of Magnetic Helicity in Solar Convective Zone: Observations and
Dynamo Theory. {\it Mon. Not. R. Astron. Soc.}, 2006, {\bf 365},
276-286.
 }

\appendix

\section{The nonlinear electromotive force in turbulent
convection}

\renewcommand{\theequation}
            {A.\arabic{equation}}

Let us derive equations for the second-order moments in a turbulent
convection. For this purpose we rewrite equations
(\ref{B1})-(\ref{B3}) in a Fourier space. In particular,
\begin{eqnarray}
&& {{\rm d}v_i({\bf k},t) \over {\rm d}t} = E_{im} \, \hat
S_{m}^{(b)}(b;{H}) + g \, e_n \, P_{in}(k) \, s({\bf k},t)
\nonumber \\
&& + \Lambda_\rho \, D_{imn}(k) \, \hat S_{mn}^{(a)}(b;{H}) + {{\rm
i}\, \Lambda_\rho \over 2 k^2} \, g \, k_m \, P_{im}(e) \, s({\bf
k},t)
\nonumber \\
&& + v_i^N \;,
\label{B4}\\
&&{{\rm d}b_i({\bf k},t) \over {\rm d}t} = {\Lambda_\rho \over 2} \,
R_{imn} \, \hat S_{mn}^{(a)}(v;{H}) + {\rm i}\, \, k_m \, \hat
S_{mi}^{(a)}(v;{H})
\nonumber \\
&& - \hat S_{i}^{(c)}(v;{H}) + b^N_i \;, \label{B5}
\end{eqnarray}
where we multiply equation~(\ref{B1}) written in ${\bf k}$-space by
$ P_{ij}({\bf k}) = \delta_{ij} - k_{ij} $ in order to exclude the
pressure term from the equation of motion, and
\begin{eqnarray*}
\hat S_{ij}^{(a)}(a; A) &=& \int a_j ({\bf k}-{\bf Q})  A_i ({\bf
Q}) \,{\rm d} {\bf Q} \;,
\\
\hat S_{i}^{(b)}(a; A) &=& (2 P_{in}(k) - \delta_{in}) \hat
S_{n}^{(c)}(a; A) + i k_n \hat S_{ni}^{(a)}(a; A) \;,
\\
\hat S_{i}^{(c)}(a; A) &=& {\rm i}\, \int a_p ({\bf k}-{\bf Q})
Q_{p} A_i ({\bf Q}) \,{\rm d} {\bf Q} \;,
\\
E_{ij} &=& \delta_{ij} - ({\rm i}\, \Lambda_\rho / k^2) (k_i e_j -
\delta_{ij} ({\bf k} {\bf \cdot} {\bf e})) \;,
\\
D_{imn} &=& e_p P_{ip}(k) \delta_{mn} + e_p k_{mp} \delta_{in} -
(1/2) e_n \delta_{im} \;,
\\
R_{imn} &=& e_m \delta_{in} - e_n \delta_{im} \;,
\end{eqnarray*}
$ \, P_{ij}(k) = \delta_{ij} - k_{ij} ,$ $ \, P_{ij}(e) =
\delta_{ij} - e_{ij} ,$ $ \, \delta_{ij} $ is the Kronecker tensor,
$ k_{ij} = k_i  k_j / k^2 $ and $ e_{ij} = e_i e_j .$ Here we
neglect terms $ \sim {\rm O}(\Lambda_\rho^2) .$ We use the
mean-field approach, and the two-point correlation function of the
velocity fluctuations is given by
\begin{eqnarray*}
\langle v_i ({\bf x}) v_j ({\bf  y}) \rangle = \int f_{ij}({\bf k,
R}) \exp{({\rm i}\, {\bf k} {\bf \cdot} {\bf r}) } \,{\rm d} {\bf
k} \;,
\end{eqnarray*}
where hereafter we omit argument $t$ in the correlation functions,
$f_{ij}({\bf k, R}) = \hat L(v_i; v_j) ,$ and
\begin{eqnarray}
\hat L(a; c) &=& \int \langle a(t,{\bf k} + {\bf  K} / 2) c(t,-{\bf
k} + {\bf  K} / 2) \rangle
\nonumber \\
&& \times \exp{({\rm i}\, {\bf K} {\bf \cdot} {\bf R}) } \,{\rm d}
{\bf  K} \;, \label{BBB2}
\end{eqnarray}
(see Roberts and Soward 1975). Here ${\bf R} = ( {\bf x} + {\bf y})
/ 2  , \quad {\bf r} = {\bf x} - {\bf y}$. Note that $ {\bf R} $ and
$ {\bf K} $ correspond to the large scales, and $ {\bf r} $ and $
{\bf k} $ to the  small ones.

Using equations~(\ref{B3}), (\ref{B4})-(\ref{B5}) we derive
equations for the following correlation functions:
\begin{eqnarray}
f_{ij}({\bf k}) &=& \hat L(v_i; v_j) \;, \quad h_{ij}({\bf k}) =
\hat L(b_i; b_j) \;,
\nonumber\\
F_{i}({\bf k}) &=& \hat L(s; v_i) \;, \quad g_{ij}({\bf k}) = \hat
L(b_i; v_j) \;,
\nonumber \\
G_{i}({\bf k}) &=& \hat L(s; b_i) \;, \quad \Theta({\bf k}) = \hat
L(s; s) \; . \label{BBB1}
\end{eqnarray}
These equations are given by
\begin{eqnarray}
{\partial f_{ij}({\bf k}) \over \partial t} &=& {\rm i}\,({\bf k}
{\bf \cdot} {\bf H}) \Phi_{ij} + I^f_{ij} + \hat{\cal N} f_{ij} \;,
\label{B6} \\
{\partial h_{ij}({\bf k}) \over \partial t} &=& - {\rm i}\,({\bf
k}{\bf \cdot} {\bf H}) \Phi_{ij} + I^h_{ij} + \hat{\cal N}h_{ij} \;,
\label{B7} \\
{\partial g_{ij}({\bf k }) \over \partial t} &=& {\rm i}\,({\bf k}
{\bf \cdot} {\bf H}) [f_{ij}({\bf k}) - h_{ij}({\bf k})]
\nonumber \\
&& + g e_n P_{jn}(k) G_{i}(-{\bf k}) + I^g_{ij} + \hat{\cal N}
g_{ij} \;,
\label{B8} \\
{\partial F_{i}({\bf k}) \over \partial t} &=& - {\rm i}\,({\bf k}
{\bf \cdot} {\bf H}) G_{i}({\bf k}) + g e_n P_{in}(k) \Theta({\bf
k})
\nonumber \\
&& + I^F_{i} + \hat{\cal N} F_{i} \;,
\label{B9} \\
{\partial G_{i}({\bf k}) \over \partial t} &=& - {\rm i}\,({\bf k}
{\bf \cdot} {\bf H}) F_{i}({\bf k}) + I^G_{i} + \hat{\cal N} G_{i}
\;,
\label{B10} \\
{\partial \Theta({\bf k}) \over \partial t} &=& - {\Omega_b^2 \over
g} F_{z}({\bf k}) + \hat{\cal N} \Theta \;, \label{B11}
\end{eqnarray}
where hereafter we also omit argument ${\bf R}$ in the correlation
functions. Here $\Phi_{ij}({\bf k }) = g_{ij}({\bf k}) -
g_{ji}(-{\bf k}) ,$ and
\begin{eqnarray*}
I^f_{ij} &=& \tilde I^f_{ij}({\bf k}) + \tilde I^f_{ji}(-{\bf k})
\;, \, I^h_{ij} = \tilde I^h_{ij}({\bf k}) + \tilde I^h_{ji}(-{\bf
k}) \;, \,
\\
\tilde I^f_{ij}({\bf k}) &=& N^f_{in} g_{nj}({\bf k}) + M_{i}
F_{j}({\bf k}) \;, \; \tilde I^h_{ij}({\bf k}) = N^h_{in}
g_{jn}(-{\bf k}) ,
\\
I^g_{ij} &=& N^h_{in} f_{nj}({\bf k}) + N^f_{jn} h_{in}({\bf k}) +
M_{j} G_{i}(-{\bf k}) \;,
\\
I^F_{i} &=& N^f_{in} G_{n}({\bf k}) - M_{i} \Theta({\bf k}) \;, \;
\; I^G_{i} = N^h_{in} F_{n}({\bf k}) \;,
\\
N^f_{ij} &=& \Lambda_\rho (D_{imj} + k_{im} e_j - \delta_{ij} k_{nm}
e_n)  H_m
\nonumber \\
&& + (2 P_{im}(k) - \delta_{im})  H_{m,j}
\\
&& + {1 \over 2} \delta_{ij} \biggl({\bf H} {\bf \cdot} \bec{\nabla}
-  H_{n,q} \, k_n \, {\partial \over \partial k_q} \biggr) \;,
\\
N^h_{ij} &=& {1 \over 2} \biggl[ \Lambda_\rho R_{imj}  H_m +
\delta_{ij} \biggl({\bf H} {\bf \cdot} \bec{\nabla} -  H_{n,q} \,
k_n \, {\partial \over \partial k_q} \biggr)\biggr]
\nonumber \\
&& -  H_{i,j} \;,
\\
M_{i} &=& {{\rm i}\, g \over 2k^2} [e_m (P_{mn}(k) k_{i} + P_{in}(k)
k_{m}) \nabla_n
\nonumber \\
&& - \Lambda_\rho P_{in}(e) k_{n}] \;,
\end{eqnarray*}
$ \bec{\nabla} = \partial / \partial {\bf R} $ and $ H_{i,j} =
\nabla_j  H_{i} ,$ $\, \hat{\cal N} f_{ij} = g e_n [P_{in}(k)
F_{j}({\bf k}) + P_{jn}(k) F_{i}(-{\bf k})] + \hat{\cal N} \tilde
f_{ij}$, and $ \hat{\cal N}\tilde f_{ij}$, $\, \hat{\cal N}h_{ij}$,
$\, \hat{\cal N}g_{ij}$, $\, \hat{\cal N}F_{i}$, $\, \hat{\cal
N}G_{i}$ and $\hat{\cal N}\Theta$ are the third-order moment terms
appearing due to the nonlinear terms. The terms $\sim F_i$ in the
tensor $ \hat{\cal N}f_{ij}$ can be considered as a stirring force
for the turbulent convection. Note that a stirring force in the
Navier-Stokes turbulence is an external parameter.

For the derivation of equations~(\ref{B6})-(\ref{B11}) we use an
identity for the function $Z_{ij}({\bf k, R})$:
\begin{eqnarray*}
&& Z_{ij}({\bf k, R}) =  {\rm i}\, \int (k_{p} + K_{p}/2) H_{p}({\bf
Q}) \exp(i {\bf K} {\bf \cdot} {\bf R})
\nonumber \\
&& \times \langle v_i ({\bf k} + {\bf K} / 2 - {\bf  Q}) v_j(-{\bf
k} + {\bf  K}  / 2)\rangle \,{\rm d} {\bf K} \,{\rm d} {\bf  Q} \; .
\end{eqnarray*}
The identity reads
\begin{eqnarray}
Z_{ij}({\bf k},{\bf R}) &\simeq& [{\rm i}\,({\bf k } {\bf \cdot}
{\bf H}) + (1/2) ({\bf H} {\bf \cdot} \bec{\nabla})] f_{ij}({\bf
k},{\bf R})
\nonumber \\
&& - \frac{1}{2} k_{p} {\partial f_{ij}({\bf k}) \over
\partial k_s}  H_{p,s} \;,
\label{B16}
\end{eqnarray}
(see Rogachevskii and Kleeorin 2004), and similarly for other
second-order moments. We take into account that in
equation~(\ref{B8}) the terms with symmetric tensors with respect to
the indexes "i" and "j" do not contribute to the nonlinear
electromotive force. In equations~(\ref{B6})-(\ref{B11}) we neglect
the second-order and high-order spatial derivatives with respect to
the large-scale variable ${\bf R}$.

Let us solve equations~(\ref{B6})-(\ref{B11}) neglecting the sources
$I^f_{ij}, I^h_{ij}, I^g_{ij}, ...$ with the large-scale spatial
derivatives. Then we take into account the terms with the
large-scale spatial derivatives by perturbations. Thus, subtracting
equations~(\ref{B6})-(\ref{B11}) written for background turbulent
convection (i.e., for ${\bf B}=0)$ from those for ${\bf B} \not=0$,
using the spectral $\tau$ approximation [which is determined by
equation~(\ref{AAC3})], neglecting the terms with the large-scale
spatial derivatives, assuming that $\eta k^2 \ll \tau^{-1}$ and $\nu
k^2 \ll \tau^{-1}$ for the inertial range of turbulent fluid flow,
and assuming that the characteristic time of variation of the mean
magnetic field ${\bf B}$ is substantially larger than the
correlation time $\tau(k)$ for all turbulence scales, we arrive to
the following steady-state solution of the obtained equations:
\begin{eqnarray}
\hat f_{ij}({\bf k}) &\approx& f_{ij}^{(0)}({\bf k}) + {\rm i}\,
\tau ({\bf k} {\bf \cdot} {\bf H}) \hat \Phi_{ij}({\bf k}) \;,
\label{B17}\\
\hat h_{ij}({\bf k}) &\approx& h_{ij}^{(0)}({\bf k}) - {\rm i}\,
\tau ({\bf k} {\bf \cdot} {\bf H}) \hat \Phi_{ij}({\bf k}) \;,
\label{B18}\\
\hat g_{ij}({\bf k}) &\approx& {\rm i}\, \tau ({\bf k} {\bf \cdot}
{\bf H}) [\hat f_{ij}({\bf k}) - \hat h_{ij}({\bf k})]
\nonumber \\
&& - \tau g e_n P_{jn}(k) \hat G_{i}({\bf k}) \;,
\label{B19}\\
\hat F_{i}({\bf k}) &\approx& F_{i}^{(0)}({\bf k}) - {\rm i}\,
\tau({\bf k} {\bf \cdot} {\bf H}) \hat G_{i}({\bf k})
\nonumber \\
&& + \tau g e_n P_{in}(k) [\hat \Theta({\bf k}) - \Theta^{(0)}({\bf
k})] \;,
\label{B20}\\
\hat G_{i}({\bf k}) &\approx& - {\rm i}\, \tau ({\bf k} {\bf \cdot}
{\bf H}) \hat F_{i}({\bf k}) \;,
\label{B21A}\\
\hat \Theta({\bf k}) &\approx& \Theta^{(0)}({\bf k}) +
O(\Omega_b^2)  \;,
\label{B21}
\end{eqnarray}
where $\hat f_{ij}, \, \hat h_{ij}, ... , \hat \Theta $ are the
solutions without the sources $I^f_{ij}, \, I^h_{ij} , ... , I^G_{i}
$ and $\hat \Phi_{ij}({\bf k}) = \hat g_{ij}({\bf k}) - \hat
g_{ji}(-{\bf k}) .$ The quantities with the superscript $(0)$ in
equations~(\ref{B17})-(\ref{B21}) correspond to the background
turbulent convection. Here we take into account that for the
background turbulent convection $g_{ij}^{(0)}({\bf k}) = 0$ and
$G_{i}^{(0)}({\bf k}) = 0$.

Now we split all second-order correlation functions into symmetric
and antisymmetric parts with respect to the wave vector ${\bf k}$,
i.e., $ f_{ij} = f_{ij}^{(s)} + f_{ij}^{(a)} ,$ where $ f_{ij}^{(s)}
= [f_{ij}({\bf k}) + f_{ij}(-{\bf k})] / 2 $ and $ f_{ij}^{(a)} =
[f_{ij}({\bf k}) - f_{ij}(-{\bf k})] / 2 .$ Thus, equations
(\ref{B17})-(\ref{B21}) yield
\begin{eqnarray}
\hat f_{ij}^{(s)}({\bf k}) &\approx& {1 \over 1 + 2 \psi} [(1 +
\psi) f_{ij}^{(0s)}({\bf k}) + \psi h_{ij}^{(0s)}({\bf k})
\nonumber \\
&& - 2 \psi \tau g e_n P_{in}(k) \hat F_{j}^{(s)}({\bf k})]  \;,
\label{B22}\\
\hat h_{ij}^{(s)}({\bf k}) &\approx& {1 \over 1 + 2 \psi} [\psi
f_{ij}^{(0s)}({\bf k}) + (1 + \psi) h_{ij}^{(0s)}({\bf k})
\nonumber \\
&& + \psi \tau g e_n P_{in}(k) \hat F_{j}^{(s)}({\bf k})] \;,
\label{B24}\\
\hat g_{ij}^{(a)}({\bf k}) &\approx& {{\rm i}\, \tau ({\bf k} {\bf
\cdot} {\bf H}) \over 1 + 2 \psi} [f_{ij}^{(0s)}({\bf k}) -
h_{ij}^{(0s)}({\bf k})
\nonumber \\
&& + \tau g e_n P_{in}(k) \hat F_{j}^{(s)}({\bf k})]  \;,
\label{B26}\\
\hat F_{i}^{(s)}({\bf k}) &\approx& {F_{i}^{(0s)}({\bf k}) \over 1
+ \psi/2} \;,
\label{B20B}\\
\hat G_{i}^{(a)}({\bf k}) &\approx& - {\rm i}\, \tau ({\bf k} {\bf
\cdot} {\bf H}) \hat F_{i}^{(s)}({\bf k}) \;, \label{B21B}
\end{eqnarray}
where $ \psi({\bf k}) = 2 (\tau \, {\bf k} {\bf \cdot} {\bf H})^2 $
and we neglect terms $\sim {\rm O}(\Omega_b^2)$ in
equations~(\ref{B21}). The correlation functions $\hat
f_{ij}^{(a)}$, $\, \hat h_{ij}^{(a)}$,$\, \hat g_{ij}^{(s)}$ $\,
\hat F_{i}^{(a)}$ and $\, \hat G_{i}^{(s)}$ vanish if we neglect the
large-scale spatial derivatives, i.e., they are proportional to the
first-order spatial derivatives.

Now we take into account the large-scale spatial derivatives in
equations (\ref{B6})-(\ref{B11}) by perturbations. Their effects
determine the following steady-state equations for the second
moments:
\begin{eqnarray}
\tilde f_{ij}^{(a)}({\bf k}) &=& f_{ij}^{(0a)}({\bf k}) + {\rm i}\,
\tau ({\bf k} {\bf \cdot} {\bf H}) \tilde \Phi_{ij}^{(s)}({\bf k})
\nonumber \\
&& + \tau I^f_{ij} \;,
\label{B29} \\
\tilde h_{ij}^{(a)}({\bf k}) &=& h_{ij}^{(0a)}({\bf k}) - {\rm i}\,
\tau ({\bf k} {\bf \cdot} {\bf H}) \tilde \Phi_{ij}^{(s)}({\bf k})
\nonumber \\
&& + \tau I^h_{ij} \;,
\label{B29B} \\
\tilde g_{ij}^{(s)}({\bf k }) &=& \tau [{\rm i}\, ({\bf k} {\bf
\cdot} {\bf H}) (\tilde f_{ij}^{(a)}({\bf k}) - \tilde
h_{ij}^{(a)}({\bf k}))
\nonumber \\
&& + g e_n P_{jn}(k) \tilde G_{i}^{(s)}(-{\bf k}) + I^g_{ij} ] \;,
\label{B28} \\
\tilde G_{i}^{(s)}({\bf k}) &=& - \tau [{\rm i}\,({\bf k} {\bf
\cdot} {\bf H}) \tilde F_{i}^{(a)}({\bf k}) - I^G_{i}] \;,
\label{B30} \\
\tilde F_{i}^{(a)}({\bf k}) &=& F_{i}^{(0a)}({\bf k}) - \tau [{\rm
i}({\bf k} {\bf \cdot} {\bf H}) \tilde G_{i}^{(s)}({\bf k})
\nonumber \\
&& - I^F_{i} ] \;, \label{B31}
\end{eqnarray}
where the second moments $\tilde f_{ij}$, $\, \tilde h_{ij}$ $\,
\tilde g_{ij}, ...$ determine the effect of the large-scale
derivatives and $\tilde \Phi_{ij}^{(s)}({\bf k}) = \tilde
g_{ij}^{(s)}({\bf k}) - \tilde g_{ji}^{(s)}(-{\bf k}) .$ The
correlation functions of the background turbulent convection
$f_{ij}^{(0a)}({\bf k})$, $h_{ij}^{(0a)}({\bf k})$ and
$F_{i}^{(0a)}({\bf k})$ are determined by the inhomogeneities of
turbulence, the fluid density and the turbulent heat flux [see
equations~(\ref{K1})-(\ref{K2}) below]. Equations~(\ref{B30}) and
(\ref{B31}) yield
\begin{eqnarray}
\tilde F_{i}^{(a)}({\bf k}) &=& {1 \over 1 + \psi/2} [F_{i}^{(0a)} -
{\rm i}\,({\bf k} {\bf \cdot} {\bf H}) \tau I^G_{i}
\nonumber \\
&& + \tau I^F_{i}] \;,
\label{B33}\\
\tilde G_{i}^{(s)}({\bf k}) &=& - {\tau \over 1 + \psi/2} [{\rm
i}({\bf k} {\bf \cdot} {\bf H}) (F_{i}^{(0a)} + \tau I^F_{i})
\nonumber \\
&& - I^G_{i}] \; . \label{B32}
\end{eqnarray}

Our goal is to calculate the mean electromotive force $ {\cal
E}_{i}({\bf r}=0) = (1/2 \sqrt{\rho_0}) \,  \varepsilon_{inm} \,
\int \tilde \Phi_{mn}^{(s)}({\bf k}) \,{\rm d} {\bf k} .$ Solution
of system of equations (\ref{B29})-(\ref{B28}) allow us to get the
expression for $\tilde \Phi_{mn}^{(s)}({\bf k})$ which yields the
mean electromotive force:
\begin{eqnarray}
{\cal E}_{i} &=& \int {\tau \varepsilon_{inm} \over 1 + 2 \psi}
\biggl[{\rm i}\, ({\bf k} {\bf \cdot} {\bf H}) [f_{mn}^{(0a)} -
h_{mn}^{(0a)} + \tau (I^f_{mn} - I^h_{mn})]
\nonumber \\
&& + I^g_{mn} + {\tau g e_p P_{mp}(k) \over 1 + \psi/2} [ {\rm
i}\,({\bf k} {\bf \cdot} {\bf H}) (F_{n}^{(0a)}
\nonumber \\
&& + \tau I^F_{n}) - I^G_{n}] \biggr] \,{\rm d} {\bf k} \;,
\label{B34}
\end{eqnarray}
where we use equations~(\ref{B33}) and (\ref{B32}).
Equation~(\ref{B34}) can be rewritten in the form:
\begin{eqnarray}
{\cal E}_{i} &=& \int {\tau \varepsilon_{inm} \over 1 + 2 \psi}
\biggl\{{\rm i}\, ({\bf k} {\bf \cdot} {\bf H}) \{f_{mn}^{(0a)} -
h_{mn}^{(0a)} + 2 \tau [(N^f_{mp}
\nonumber \\
&& + N^h_{mp}) \hat g_{pn} - M_n (\hat F_{m} - F_{m}^{(0s)})]\} +
N^h_{mp} \hat f_{pn}
\nonumber \\
&& - N^f_{mp} \hat h_{pn} + {\rm i}\, \tau M_n ({\bf k} {\bf \cdot}
{\bf H}) \hat F_{m} + {\tau g e_p P_{mp}(k) \over 1 + \psi/2}
\nonumber \\
&& \times \biggl[{\rm i}\,({\bf k} {\bf \cdot} {\bf H}) F_{n}^{(0a)}
+ \biggl({\psi \over 2} N^f_{nq} - N^h_{nq} \biggr) \hat F_{q}
\biggr] \biggr\} \,{\rm d} {\bf k} \; .
\nonumber \\
\label{B34B}
\end{eqnarray}
For the integration in $ {\bf k} $-space in equation~(\ref{B34B}) we
specify a model for the background turbulent convection (i.e., the
turbulence with zero mean magnetic field, $ {\bf B} = 0)$, which is
determined by
\begin{eqnarray}
f_{ij}^{(0)}({\bf k}) &=& f_{\ast} \, W(k) \, \biggl[P_{ij}({\bf k})
+ {{\rm i}\, \over 2 k^2} \, (k_i \Lambda_j^{(v)}
\nonumber \\
&& - k_j \Lambda_i^{(v)})\biggr] \;,
\label{K1} \\
h_{ij}^{(0)}({\bf k}) &=& h_{\ast} \, W(k) \, \biggl[P_{ij}({\bf k})
+ {{\rm i}\, \over 2 k^2} \, (k_i \Lambda_j^{(b)}
\nonumber \\
&& - k_j \Lambda_i^{(b)})\biggr] \;,
\label{KK1} \\
F^{(0)}_{i}({\bf k}) &=& 3 \, F_{\ast} \,  W(k) \, e_j \,
\biggl[P_{ij}({\bf k}) - {{\rm i}\, \over 2 k^2} \, (P_{jm}({\bf k})
k_{i}
\nonumber \\
&& + P_{im}({\bf k}) k_{j}) \tilde \Lambda_m^{(F)}\biggr] \;,
\label{K2}
\end{eqnarray}
$\Theta^{(0)}({\bf k}) =  2 \, \Theta_{\ast} \, W(k)$, $\,
g_{ij}^{(0)}({\bf k}) = 0$ and $G_{i}^{(0)}({\bf k}) = 0$, where
$P_{ij}({\bf k}) = \delta_{ij} - k_{ij}$, $\, k_{ij} = k_{i} k_{j} /
k^{2}$, $\, W(k) = E(k) / 8 \pi k^{2}$, $\, \tau(k) = 2 \, \tau_{0}
\, \bar \tau(k)$, $\, E(k) = - d \bar \tau(k) / dk$, $\, \bar
\tau(k) = (k / k_{0})^{1-q}$, $\, 1 < q < 3$ is the exponent of the
kinetic energy spectrum (e.g., $q = 5/3$ for Kolmogorov spectrum),
$k_{0} = 1 / l_{0}$ and $\tau_{0} = l_{0} / u_{0}$. Here
$\Lambda_i^{(v)} = \Lambda_i^{(u)} - 2 \Lambda_\rho e_i$ and $\tilde
\Lambda_i^{(F)} = \Lambda_i^{(F)} - 2 \Lambda_\rho e_i$. These imply
that
\begin{eqnarray*}
\Lambda_i^{(v)} = {\nabla_i (\rho_0^2 \, \langle {\bf u}'^2
\rangle^{(0)}) \over \rho_0^2 \, \langle {\bf u}'^2 \rangle^{(0)}}
\;, \quad  \quad \quad \tilde \Lambda_i^{(F)} = {\nabla_i (\rho_0^2
\, \langle |{\bf u}'| \, s' \rangle^{(0)}) \over \rho_0^2 \, \langle
|{\bf u}'| \, s' \rangle^{(0)}} \;,
\end{eqnarray*}
where
\begin{eqnarray*}
&& \Lambda_i^{(u)} = {\nabla_i \langle {\bf u}'^2 \rangle^{(0)}
\over \langle {\bf u}'^2 \rangle^{(0)}} \;, \quad  \quad \quad
\Lambda_i^{(b)} = {\nabla_i \langle {\bf b}^2 \rangle^{(0)} \over
\langle {\bf b}^2 \rangle^{(0)}} \;,
\\
&& \Lambda_i^{(F)} = {\nabla_i \langle |{\bf u}'| \, s'
\rangle^{(0)} \over \langle |{\bf u}'| \, s' \rangle^{(0)}}
\end{eqnarray*}
and $\int F^{(0)}_i({\bf k}) \,{\rm d} {\bf k} = F_{\ast} \, e_i$,
$\, \int f_{ij}^{(0)}({\bf k}) \,{\rm d} {\bf k} = (f_{\ast} / 3) \,
\delta_{ij}$, $\, \int h_{ij}^{(0)}({\bf k}) \,{\rm d} {\bf k} =
(h_{\ast} / 3) \, \delta_{ij}$ and $\, \int \Theta^{(0)}({\bf k})
\,{\rm d} {\bf k} = \Theta_{\ast}$.

After the integration in ${\bf k}$ space in equation~(\ref{B34B}) we
obtain the nonlinear electromotive force.  This yields the nonlinear
turbulent magnetic diffusion coefficients and the nonlinear drift
velocities of the mean magnetic field in the axisymmetric case,
which are given by equation~(\ref{C9}), where the contributions from
the purely hydrodynamic turbulence are given by
\begin{eqnarray}
\eta_{_{A}}^{(v)}({B}) &=& A_{1}^{(1)}(4  B) + A_{2}^{(1)}(4  B) \;,
\label{AA1} \\
\eta_{_{B}}^{(v)}({B}) &=& A_{1}^{(1)}(4  B) + 3 (1 - \epsilon)
\biggl[A_{2}^{(1)}(4  B)
\nonumber \\
&& - {1 \over 2\pi} \bar A_{2}(16  B^2)\biggr] \;,
\label{AA2} \\
{\bf V}_{A}^{(v)}({B}) &=& - {1 \over 2} \eta_{_{A}}^{(v)}({B})
({\bf \Lambda}^{(u)} - \epsilon {\bf \Lambda}^{(b)}) + {\bf
V}^{(u,\rho)}
\nonumber \\
&& - {{\bf \Lambda}^{(B)} \over 2} \biggl[(2 - 3 \epsilon)
A_{2}^{(1)}(4  B)
\nonumber \\
&& -  {3 (1 - \epsilon)\over 2 \pi} \bar A_{2}(16 B^2)\biggr] ,
\label{AA3} \\
{\bf V}_{B}^{(v)}({B}) &=& - {1 \over 2} \eta_{_{A}}^{(v)}({B})
({\bf \Lambda}^{(u)} - \epsilon {\bf \Lambda}^{(b)})
\nonumber \\
&& + {\bf V}^{(u,\rho)} \;,
\label{AA4} \\
{\bf V}^{(u,\rho)} &=& {1 \over 2} \Lambda_\rho {\bf e}
\biggl[\epsilon A_{1}^{(1)}(4  B) - (5 - 6\epsilon) A_{2}^{(1)}(4 B)
\nonumber \\
&& + {3 (1 - \epsilon)\over 2 \pi} \bar A_{2}(16  B^2) \biggr] \;,
\label{AA5}
\end{eqnarray}
and the contributions caused by the turbulent heat flux are
\begin{eqnarray}
\eta_{_{A}}^{(F,z)}({B}) &=& {3 \over 4} \biggl[2 \Psi_1\{A_{2} - 3
A_{1} + 3 C_{1} \} + 4 \Psi_2 \{A_{1} - C_{1}\}
\nonumber \\
&& + 3 \Psi_3 \{A_{1} + C_{1}\} \biggr] \;,
\label{AA6} \\
\eta_{_{A}}^{(F,x)}({B}) &=& {3 \over 4} \biggl[ - 2 \Psi_1 \{A_{1}
+ C_{1}\} + 4 \Psi_2\{C_{1}\}
\nonumber \\
&& + 3 \Psi_3 \{A_{1} - 2 C_{1}\} \biggr] \;,
\label{AA7} \\
\eta_{_{B}}^{(F,z)}({B}) &=& {3 \over 4} \biggl[ (-6\Psi_1 + 4
\Psi_2 + 3 \Psi_3) \{A_{1} + A_{2} - C_{1}
\nonumber \\
&& - C_{3} \} + 2\Psi_1\{C_{3}\} + 6\Psi_3\{C_{1}\} \biggr] \;,
\nonumber \\
\label{AA8} \\
\eta_{_{B}}^{(F,x)}({B}) &=& 0  \;,
\label{AA9} \\
{\bf V}_{A}^{(F)}({B}) &=& {\bf V}_{B}^{(F)}({B}) = {\bf V}^{(F)} +
{\bf V}^{(F,\rho)}\;,
\label{AA10} \\
V^{(F)}_z({B}) &=& - {3 \over 4} \Lambda_z^{(F)} \biggl[\Psi_4
\{A_{1} + A_{2} - C_{1} - C_{3} \}
\nonumber \\
&& - A_{1}^{(2)}(4  B) - A_{2}^{(2)}(4  B) + C_{1}^{(2)}(4  B)
\nonumber \\
&& + C_{3}^{(2)}(4  B)\biggr] \;,
\label{AA11} \\
V^{(F)}_x({B}) &=& {3 \over 4} \Lambda_x^{(F)} \biggl[\Psi_4 \{A_{1}
+ A_{2} - 5 C_{1} - 5 C_{3} \}
\nonumber \\
&& - A_{1}^{(2)}(4  B) - A_{2}^{(2)}(4 B) + 5 C_{1}^{(2)}(4  B)
\nonumber \\
&& + 5 C_{3}^{(2)}(4 B)\biggr] \;,
\label{AA12} \\
{\bf V}^{(F,\rho)} &=& {3 \over 8} \Lambda_\rho {\bf e}
\biggl[\Psi_1\{17 A_{1} + 17 A_{2} + 17 C_{1} - 7 C_{3}\}
\nonumber \\
&& - \Psi_2 \{6 A_{1} + 6 A_{2} + 10 C_{1} - 2 C_{3}\} - \Psi_3 \{9
A_{1}
\nonumber \\
&& + 9 A_{2} + 15 C_{1} - 3 C_{3}\} - 4 A_{1}^{(2)}(4  B)
\nonumber \\
&& - 4 A_{2}^{(2)}(4  B)\biggr] \;, \label{AA14}
\end{eqnarray}
where ${\bf \Lambda}^{(B)} = (\bec{\nabla} {\bf B}^2) / {\bf B}^2$,
the parameter $\epsilon = \langle {\bf b}^2 \rangle^{(0)} / \langle
{\bf v}^2 \rangle^{(0)}$, the parameter $a_\ast$ is given by
equation~(\ref{AAC1}), and
\begin{eqnarray}
\Psi_1\{X\} &=& {1 \over 3} [4 X^{(2)}(4  B) - X^{(2)}(2  B)] \;,
\nonumber \\
\Psi_2\{X\} &=& {1 \over 9}\biggl[16 X^{(2)}(4  B) - 16 X^{(2)}(2 B)
\nonumber \\
&& + {9 \over 2 \pi}  \bar X(4  B^2)\biggr]\;,
\nonumber \\
\Psi_3\{X\} &=& - {1 \over 9} \biggl[4 X^{(2)}(4  B) - 49 X^{(2)}(2
B)
\nonumber \\
&& + {18 \over \pi} \bar X(4  B^2 )\biggr] \;,
\nonumber \\
\Psi_4 \{X\} &=& 3 X^{(1)}(4  B) - {3 \over 2 \pi} \bar X(16  B^2)
\; . \label{B38}
\end{eqnarray}
Note that $\Psi_4\{A_{1}\} = A_{1}^{(1)} + (1/2) A_{2}^{(1)} .$ The
functions $ A_{m}^{(n)}(\beta) $, $\, C_{m}^{(n)}(\beta) $ for $n =
1; \, 2$ and the functions $\bar A_{m}(\beta^2) $, $\, \bar
C_{m}(\beta^2) $ are given in Rogachevskii and Kleeorin (2004, in
Appendixes B, C and D). Asymptotic formulae for the nonlinear
turbulent magnetic diffusion coefficients and the nonlinear drift
velocities of the mean magnetic field in the axisymmetric case are
given by equations~(\ref{AAC9}) and~(\ref{AAC10}).

\end{document}